\documentclass[11pt,twoside]{elsart}
\usepackage{natbib}
\usepackage{amsmath}
\usepackage{amsthm}
\usepackage{amscd}
\usepackage{amsfonts}
\usepackage{amssymb}
\def\urlprefix\url#1{\url{#1}} 
\providecommand{\href}[2]{{#2}}
\providecommand{\url}[1]{\texttt{#1}}

\newtheorem{theorem}{Theorem}
\newtheorem{propo}{Proposition}
\newtheorem{lemma}{Lemma}
\newtheorem{corol}{Corollary}
\theoremstyle{definition}
\newtheorem{defi}{Definition}
\theoremstyle{remark}
\newtheorem{remark}{Remark}

\newcommand{\C}{{\mathbb{C}}}
\newcommand{\R}{{\mathbb{R}}}
\newcommand{\Z}{{\mathbb{Z}}}
\newcommand{\CS}{$C^*$}

\newcommand{\hookllongrightarrow}{\lhook\joinrel\relbar\joinrel\longrightarrow}

\DeclareMathOperator{\curv}{curv}
\DeclareMathOperator{\id}{id}
\DeclareMathOperator{\Hom}{Hom}

\begin{document}
\begin{frontmatter}
\title{Bloch Theory and Quantization of Magnetic Systems}
\author{Michael J.\ Gruber}
\address{Dept.\ of Mathematics, MIT 2--167, 77 Mass.\ Av., Cambridge, MA 02139, USA\\
\textup{\texttt{mjg@math.mit.edu}}\\
\textup{\texttt{http://www--math.mit.edu/$\tilde{}$ mjg}} 
}

\journal{J.\ Geom.\ Phys.}
\begin{abstract}
Quantizing the motion of particles on a Riemannian manifold in the presence of a magnetic field poses the problems of existence and uniqueness of quantizations. 
Both of them are considered since the early days of geometric quantization but there is still some structural insight to gain from spectral theory. 
Following the work of \cite*{AscOveSei:MBABL} for the 2--torus we describe the relation between quantization on the manifold and Bloch theory on its covering space for more general compact manifolds.

\end{abstract}
\begin{keyword}
Geometric quantization\sep
spectral theory\sep
Bloch theory\sep
Bochner Laplacian\sep
Schr\"odinger operator\sep
magnetic fields
\MSC 81S10 (primary); 58F06, 58G25, 81Q10 (secondary)
\PACS 02.40.Vh (primary); 03.65, 02.30 (secondary)
\end{keyword}
\end{frontmatter}


\section*{Introduction}
In geometric quantization for symplectic manifolds one is faced with questions of existence and uniqueness \cite[see e.g.\ ][]{SimWoo:LGQ,Woo:GQ} which do not arise for the common phase space $T^*M$ (with standard symplectic structure) of Hamiltonian mechanics.
But, when incorporating magnetic fields (closed 2-forms $b\in\Omega^2(M)$) into the picture one is forced either to choose magnetic potentials ($a\in\Omega^1(M)$ with $da=b$) or to ``charge'' the standard symplectic structure by the magnetic field
(see remark \ref{remark:minimal coupling} below).
In either case, the questions of existence and uniqueness come up now even for the phase space $T^*M$.
Indeed, these questions arise for prequantizations, whereas --- given a prequantization --- there is a canonical choice of a quantization  when the phace space is $T^*M$ with a charged symplectic structure (at least for Hamiltonians linear in the momenta; see remark~\ref{remark:GCBL}).

On the other hand, the cohomological obstructions and degrees of freedom for geometric quantization vanish on the covering space $X:=\tilde M$.
Since the classical Hamiltonian system may be lifted from $M$ to $X$ one may try to quantize on $X$ and push the quantization down to $M$ again.
This push down is possible if and only if the system on $M$ is quantizable.
But quantization on $X$ is unique, so one may ask which quantizations on $M$ one gets by this procedure, and how to recover the other quantizations on $M$ from that on $X$.
Since the magnetic Schr\"odinger operator $H$ arising from a quantization on $X$ is periodic (any operator arising from a periodic classical symbol is) one can, in the case of abelian covering group, analyze it using Bloch theory.
This gives a decompostion of $H$ into a direct integral of operators (the ``fibers'' of $H$) acting on line bundles over $M$.  
It turns out that the fibers are unitarily equivalent to magnetic Schr\"odinger operators arising from quantizations on $M$, and that the direct integral runs just over all classes of quantizations on $M$, using a natural integration measure.
This follows the ideas of \cite*{AscOveSei:MBABL} who did the same work for the 2-torus.

\subsection*{Outline}

In section \ref{sec:ECQ} we recall the definitions (quantization of a system with magnetic field, equivalence of quantizations) and the appropriate cohomology groups.
All of that is known from the standard literature on geometric quantization, so we will not give references to the results individually.

In section \ref{sec:C} we describe the connections between sets of equivalence classes of quantizations, as determined in the previous section, and representions of the fundamental group.

In section \ref{sec:BTVB} we recall Bloch theory in the geometric context of periodic operators acting on sections of vector bundles.

In section \ref{sec:PMF} we analyze the Bloch decomposition for Schr\"odinger operators with magnetic fields, identify the fibers of this decomposition (theorem \ref{theorem:direct integral}) and draw our final conclusions about the relation to quantization (corollary \ref{corol:Bloch theory and quantization}).

\subsection*{Acknowledgements} 

This work is a (commutative) part of my Ph.D.\ thesis ``Nichtkommutative Blochtheorie'' \cite[non-commutative Bloch theory;][]{Gru:NB}. 
I gratefully appreciate the advice and supervision given by Jochen Br\"uning at Humboldt-University at Berlin. 
\nocite{Gru:NBTO}

This work was supported by Deutsche Forschungsgemeinschaft as project D6 at the Sonderforschungsbereich 288 (differential geometry and quantum physics), where this article is available as preprint 375.

Finally I would like to thank the referee for valuable remarks on the organisation of the paper and on some pecularities of the quantization of quadratic Hamiltonians.

\section{Equivalence classes of quantizations}\label{sec:ECQ}
\begin{remark}[minimal coupling]\label{remark:minimal coupling}
 Lorentz force is described in Newton's equations of classical mechanics using a magnetic field $b\in C^\infty(T\R^3)\simeq \Omega^2(\R^3)$ (``axial vector field''). 
When trying to incorporate it into the formalism of  Lagrange or Hamilton mechanics, one is faced with the necessity (or, at least, utility) of introducing a vector potential $a\in C^\infty(T\R^3)\simeq \Omega^1(\R^3)$ (``polar vector field'') such that
$b=da$ ($b$ is divergence free, i.e.\ closed; since $H^2_{dR}(\R^3)=0$   $b$  is exact).
A Hamiltonian $h\in C^\infty(T\R^3)$ is replaced by $h_a\colon(x,p)\mapsto h(x,p-qa(x))$ (electric charge  $q$), the so--called \emph{mimimally coupled Hamiltonian}.
Doing this for a free particle ($h(x,p)=\frac{1}{2m}|p|^2$, mass $m$) one gets $h_a(x,p)=\frac{1}{2m}|p-qa(x)|^2$ which suggests using $H_a=\frac{1}{2m}\left(\frac{\hbar}{\imath}\nabla-a\right)^2$ as Hamiltonian in quantum mechanics, where $\nabla$ denotes gradient in $\R^3$.
$\nabla-\frac{\imath}{\hbar}a$ may be viewed as connection on the trivial complex line bundle $\R^3\times\C$.
Note especially that the curvature is given by $\curv(\nabla-\frac{\imath}{\hbar}a)=\frac{1}{\hbar}da=\frac{1}{\hbar}b$ if we identify the Lie algebra of $U(1)$ with $\R$ in a suitable manner ($-\imath v\mapsto v\in\R$).

In the case of non-exact magnetic fields (on a manifold $M$ with non-trivial $H_{dR}^2(M)$) one can, in general, only find local vector potentials  and local connections on locally trivial complex line bundles.
If everything fits together ``nicely'' one gets a global connection on a (global) complex line bundle with curvature $\frac{1}{\hbar}b$.
This motivates definition \ref{defi:QMF}.

Another aspect of definition \ref{defi:QMF} is given by the point of view of geometric quantization.
It rests on the observation that Hamiltonian mechanics with a (closed) magnetic field $b\in \Omega^2(M)$ can be formulated without any magnetic vector potential if one uses a ``charged'' symplectic form $\omega_b=\omega+\tilde b$ on $N:=T^*M$, where $\omega$ is the canonical symplectic form on  $T^*M$ and $\tilde b$ the pull-back $\pi^*b$ of $b$ from $M$ to $T^*M$ by the projection $\pi\colon T^*M\rightarrow M$ onto the base points.
A prequantization of such a system is given by a Hermitian line bundle $\tilde L$ over $T^*M$ with connection (covariant derivative) $\tilde\nabla$ such that $\hbar\curv(\tilde\nabla)$=$\omega_b$.
A quantization is a prequantization together with a complex polarization $P$ of $N$.
A complex polarization of $N=T^*M$ is a complex distribution (i.e.\ a family $(P_x)_{x\in N}$ of complex subspaces of the complexified tangent space $TN_\C$, locally defined by smooth frames) with the following properties:
\begin{enumerate}
\item Every $P_x$ is Lagrangian with respect to the complexified symplectic structure.
\item $\dim P\cap\bar P\cap TN$ is constant on $N$.
\item $P$ is integrable,  i.e.\ closed with respect to Lie brackets.
\end{enumerate}
Since our symplectic manifold is a cotangent space with (vertically) charged symplectic form there is a canonical polarization given by the fibration over $M$ with fiber $(T_x M)_\C,x\in M$, the \emph{vertical polarization}. To be definite: the corresponding distribution is  $V_P=\left(\ker T\pi\right)_\C$.
Polarized sections in $\tilde L$ with respect to this polarization can be viewed as sections into a complex line bundle $L$ over $M$ with $\pi^*L=\tilde L$.
Such $L$ exist because the fibers of $\pi\colon T^*M\rightarrow M$ are contractible;
 $L$ can be constructed as pull-back by the 0-section  in $T^*M$.
Finally, $\tilde\nabla$ induces a connection $\nabla$ on $L$ with curvature $b$. 
\end{remark}

\begin{remark}[geometric quantization and Bochner Laplacians]\label{remark:GCBL}
In general, geometric quantization provides for means to quantize classical observables whose associated Hamiltonian flow preserves the chosen polarization.
 In the case of a cotangent space $T^*M$ with the vertical polarization mentioned above, this restricts quantization to Hamiltonians linear in the momenta in general.
There are several methods to overcome this.

Either one searches for polarizations which are invariant under the given flow. 
This has been considered especially for the geodesic flow on spheres \citep{Ii:MSQMS} and the Kepler problem \citep{Sim:GQELKP,Raw:NPPKP}.

Or one uses the Blattner-Kostant-Sternberg pairing for polarizations \citep{Bla:QRT,Kos:SS,GuiSte:GA,Sim:SEGQ,Emm:EEIQONPVP}. 
Here one may produce non-symmetric operators in general.

A third approach --- leaving the setting of geometric quantization --- consists of mimicking the Euclidean Weyl quantization (or other orderings), using normal coordinates \citep{Und:QMC,LiuQui:GIQRM,Lan:MTCQM,Pfl:DTAWQRM}.
The results depend on the choice of ordering (Weyl, normal, antinormal), wave functions (functions or half-densities) and even ones Euclidean point of view (dilations may introduce curvature terms).

In any case, the free particle Hamiltonian given by a Riemannian metric is quantized to $\Delta+\alpha R$, where we choose the convention $\Delta\geq0$, $R$ denotes scalar curvature, $\alpha$ is rational and non-negative. 
Even path integral methods and Maslov quantization lead to the same type of operator.
In physics, the Laplacian is accepted as the quantization of the free particle as well as the Bochner-Laplacian is for the particle in a magnetic field.

Since we intend to include a smooth potential $V$ in the Schr\"odinger operator anyway, one may cover any scalar curvature terms arising from some choice of quantization.
To be more specific: In section \ref{sec:PMF} we deal with periodic potentials and magnetic fields.
Since we demand the metric to be periodic also, any curvature term will be so and will simply descend to the quotient. 
Therefore, theorem~\ref{theorem:direct integral} and corollary~\ref{corol:Bloch theory and quantization} hold for any consistent choice of quantization (i.e.\ choosing $\alpha$ the same on covering and quotient), not only for the choice $\alpha=0$ made in definition~\ref{defi:QMF}.

\end{remark}

In the sequel we choose units with $\hbar=1,q=1,2m=1$.

\begin{defi}[quantization with magnetic field]\label{defi:QMF}
Let $(M,g)$ be an orientable Riemannian manifold, $b\in\Omega^2(M)$ a closed real-valued 2-form (the \emph{magnetic field}).
A \emph{quantization} of the particle motion on  $(M,g)$ in the presence of the magntic field  $b$ is given by a Hermitian line bundle $(L,h,\nabla)$ over $M$ with connection such that $\curv(\nabla)=b$.
The \emph{magnetic Schr\"odinger operator} is defined by the \emph{{Bochner-Laplacian}} 
\begin{equation}
 H^{L,\nabla}:= \nabla^\dagger\nabla\text{ with domain } {\mathcal D}(H^{L,\nabla})=C^\infty_0(L)
\end{equation}
in the Hilbert space $L^2(L)$ of square-integrable sections of $L$, defined by $g$ and $h$.  
Here, $\nabla^\dagger$ is the formal adjoint of $\nabla$.
\end{defi}

\begin{remark}[self-adjointness]
Since $H^{L,\nabla}$ is symmetric and bounded below (by 0) there is a canonical self-adjoint extension given by the Friedrichs extension $H^{L,\nabla}_F$. 
It is the self-adjoint operator associated to the closure of the symmetric form \[q(f,g):=(f,H^{L,\nabla}g)=(\nabla f,\nabla,g)\] with (form) domain ${\mathcal Q}(q)={\mathcal D}(H^{L,\nabla})$.
\end{remark}

\begin{remark}[equivalence classes of line bundles]\label{remark:ECLB}
Denote by $\underline G _M$ the sheaf of germs of smooth $G$-valued functions on $M$ for any abelian Lie group $G$.
Every complex line bundle $L$ over $M$ is defined by a \v{C}ech cocycle $(c_{\alpha\beta})\in \Check{Z}^1(M,\underline{\C^\times} _M)$.
Given any $(l_{\alpha\beta})\in \Check{C}^1(M,\underline \C _M)$ with $\exp 2\pi \imath l_{\alpha\beta}=c_{\alpha\beta}$ one has $\delta l\in \Check{Z}^2(M,\underline \Z _M)=\Check{Z}^2(M,\Z)$. Here $\delta$ denotes \v{C}ech codifferential. Other choices  $l'$ fulfil $l'-l\in \Check{C}^1(M,\Z)$, so that $\delta l$ and $\delta l'$ define the same class in $\Check{H}^2(M,\Z)$, and the mapping \[j\colon H^1(M,\underline{\C^\times}_M)\rightarrow H^2(M,\Z), c\mapsto \delta l\] is well-defined.

Every line bundle isomorphism from $c$ to $c'$ corresponds to a \v{C}ech cochain $(f_\alpha)\in \Check{C}^0(M,\underline{\C^\times}_M)$,  $c'=c\,\delta f$.

$H^2(M,\Z)$ parametrizes the set of equivalence classes of complex line bundles: The short exact sequence of sheaves
\begin{gather}
0 \xrightarrow{} \underline  \Z \xrightarrow{} \underline \C_M \xrightarrow{\exp 2\pi\imath\cdot} \underline{\C^\times}_M \xrightarrow{} 0,
\end{gather}
where
\[ \exp 2\pi\imath\cdot:\C\ni z\mapsto \exp (2\pi\imath z)\in\C^\times, 
\]
induces the following long exact sequence in \v{C}ech cohomology:
\begin{equation}
\begin{aligned}
0 & \xrightarrow{} &  H^0(M&,\underline\Z_M) &  & \xrightarrow{} &  H^0(M&,\underline\C_M) &  & \xrightarrow{} &  H^0(M&,\underline{\C^\times}_M) &  &{\xrightarrow{}} \\
  & \xrightarrow{} &  H^1(M&,\underline\Z_M) &  & \xrightarrow{} &  H^1(M&,\underline\C_M) &  & \xrightarrow{} &  H^1(M&,\underline{\C^\times}_M) &  &{\xrightarrow{j}} \\
  &      &       &\|               &  &      &       &\|  \\
  &      &  H^1(M&,\Z)             &  &      &       &0   \\
  & \xrightarrow{j} &  H^2(M&,\underline\Z_M) &  & \xrightarrow{} &  H^2(M&,\underline\C_M) &  & \xrightarrow{} & \ldots \\
  &      &       &\|               &  &      &       &\|  \\
  &      &  H^2(M&,\Z)             &  &      &       &0    
\end{aligned}
\end{equation}
So $H^i(M,\underline{\C^\times}_M)\stackrel j\simeq H^{i+1}(M,\Z)$ for every $i\geq1$, and the joining homomorphism $j$ is just the mapping described before. The class in $H^2(M,\Z)$ characterizing $L$ is called the \emph{first Chern class} $c_1(L)$ of $L$.

Every Hermitian line bundle $(L,h)$ is defined by a $(c_{\alpha\beta})\in \Check{Z}^1(M,\underline{S^1}_M)$, every Hermitian line bundle isomorphism (i.e.\ every isometry) by some $(f_\alpha)\in \Check{C}^0(M,\underline{S^1}_M)$, $c'=c\,\delta f$. Using the short exact sequence
\begin{equation}
0 \xrightarrow{} \underline  \Z \xrightarrow{} \underline \R_M \xrightarrow{ {\exp 2\pi\imath\cdot}} \underline{S^1}_M \xrightarrow{} 0
\end{equation}
and the corresponding long exact sequence in \v{C}ech cohomology one gets again $H^i(M,\underline{S^1}_M)\stackrel j\simeq H^{i+1}(M,\Z)$ for $i\geq1$, and $j$ comes from the mapping $\delta\circ\frac{\log\cdot}{2\pi\imath}$ on cochains as before.

Finally we recall that the group structure induced on $H^1(M,\underline{\C^\times}_M)$ and $H^1(M,\underline{S^1}_M)$ by the coefficient groups is just the tensor product of line bundles.
\end{remark}

\begin{remark}[integral de Rham class]
The short exact sequence of groups
\begin{equation}
0 \xrightarrow{} \Z \xrightarrow{i} \R \xrightarrow{ {\exp 2\pi\imath\cdot}} S^1 \xrightarrow{} 0
\end{equation}
induces the long exact sequence of cohomology groups
\begin{equation}
\begin{aligned}
0 & \xrightarrow{} &  H^0(M&,\Z) &  & \xrightarrow{H^0(i)} &  H^0(M&,\R) &  & \xrightarrow{} &  H^0(M&,S^1) &  &{\xrightarrow{}} \\
  &      &       &                 &  &      &       &                 &  &      &       &                          &   &{\searrow}  \\
  &      &       &                 &  &      &       &                 &  &      &       &                          &   &\phantom{\searrow}\;0  \\
  & \xrightarrow{} &  H^1(M&,\Z) &  & \xrightarrow{H^1(i)} &  H^1(M&,\R) &  & \xrightarrow{} &  H^1(M&,S^1) &  &{\xrightarrow{}} \\
  & {\nearrow} &       &                &  &      &       &  \\
0 &      &       &                 &  &      &       &   \\
  & \xrightarrow{i} &  H^2(M&,\Z) &  & \xrightarrow{H^2(i)} &  H^2(M&,\R) &  & \xrightarrow{} & \ldots \\
  &      &       &                 &  &      &       &  \\
  &      &       &                 &  &      &       &    
\end{aligned} \label{equ:exSequZRS1}
\end{equation}
A de Rham class is called  \emph{integral} if it is contained in the range of $H^*(i)$.
\end{remark}

\begin{remark}[curvature and Chern class]\label{remark:CCC}
For every line bundle with connection one has $H^*(i)(c_1(L))=[-\frac{1}{2\pi}\curv(\nabla)]$, using the identifaction $-\imath\R\simeq\R$ as in the introduction. 
This can be seen for example using Deligne cohomology with coefficients in $\R(2):=(2\pi\imath)^2\R$  \cite[see][Chap.\ 1 for these notions]{Bry:LSCCGQ}:
 Let $\mu=\delta\left(\frac{\log c}{2\pi\imath}\right)\in \check Z^2(M,\Z)$ as in remark \ref{remark:ECLB} represent $c_1(L)$ for some choice of  logarithms $\log_{\alpha\beta}$. 
This defines a cocycle in $\check Z^2(M,\R(2)^\infty_D)$ given by $(-(2\pi\imath)^2\mu, -2\pi\imath\log c,-2\pi a)$, and from a proposition on  Deligne cohomology groups $H^p(M,\R(p)^\infty_D)$ (ibidem, Lemma 1.5.4) one gets $-(2\pi\imath)^2H^*(i)([\mu])=-2\pi[da]\in H^2(M,\R)$ using the \v{C}ech--de Rham isomorphism.

This connection between curvature and Chern class immmediately implies
\end{remark}

\begin{theorem}[existence of quantizations]\label{theorem:EQ}
A system with magnetic field $(M,g,b)$ is quantizable if and only if the de Rham class of $\frac{1}{2\pi}b$ is integral.
\end{theorem}

\begin{defi}[equivalence of quantizations]\label{defi:equivalence of quantizations}
Two quantizations given by $(L,h,\nabla)$ and $(L',h',\nabla')$ are called \emph{equivalent} if there is a Hermitian line bundle isomorphism $\Phi\colon L\rightarrow L'$ intertwining the connections:
\begin{equation}
\forall{s\in C^\infty(L)}:\forall{X\in C^\infty(TM)}: \Phi\circ\nabla\!_Xs=\nabla'\!_X(\Phi \circ s) \label{equ:EQ}
\end{equation}
\end{defi}

\begin{remark}[unitary equivalence]
If $(L,h,\nabla)$ and $(L',h',\nabla')$ are two quantizations equivalent via $\Phi$, then
\begin{align*}
U_\Phi: L^2(L) &\rightarrow L^2(L'), \\
         s&\mapsto U_\Phi s := \Phi\circ s,
\end{align*}
defines a unitary operator intertwining the magnetic Schr\"odinger operators:
\[ U_\Phi H^{L,\nabla}= H^{L',\nabla'}U_\Phi \]
Conversely, if $U_\Phi$ is unitary then $\Phi$ is a Hermitian isomorphism.
Equation \ref{equ:EQ} is just the intertwining property for first order operators defined as quantizations of vector fields.
\end{remark}

\begin{remark}[local form of the gauge]
We choose a cochain $f\in\check{C}^0(M,\underline{S^1}_M)$ representing the isomorphism $\Phi$,  i.e.\ $\varphi_\alpha'\circ\Phi\circ\varphi_\alpha{}^{-1}=\id_M\times f_\alpha$, and cocycles $(c,a)$ and $(c',a')$ for $(L,h,\nabla)$ and $(L',h',\nabla')$  with respect to bundle charts $\varphi_\alpha\colon L|_{U_\alpha}\rightarrow U_\alpha\times \C$ and $\varphi_\alpha'\colon L'|_{U_\alpha}\rightarrow U_\alpha\times \C$. 
Then one easily calculates
\begin{align}
  \imath(a'_\alpha-a_\alpha) &= f_\alpha^{-1}df_\alpha=d\log f_\alpha . \label{equ:a'-a=fdf}
\end{align}
\end{remark}

\begin{remark}[2-term complex]
The ``second half'' of the condition for the De\-ligne cocycle in remark \ref{remark:CCC}, i.e.\ $-\imath(\delta a)_{\alpha\beta}=-d\log c_{\alpha\beta}$, can be viewed as  cochain condition in the 2-term complex of sheaves
\begin{equation}
 K := \left\{ \begin{CD} K^0:= \underline{S^1}_M \\ @VV{\imath d\log\cdot}V \\ K^1:=\underline{\Omega^1}_M \end{CD}  \right. \label{equ:2TKK}
\end{equation}
\end{remark}
Here $\underline{\Omega^1}_M$ denotes the sheaf of (real-valued) 1-forms on $M$. 
$(c,-a)$ defines a cocycle, hence it defines a class in the hypercohomology $H^1(M,K)$ of $K$; in \cite[][chapter 2]{Bry:LSCCGQ} it is shown that this class does not depend on the choice of line bundle isomorphism $c_{\alpha\beta}$ and connection forms $a_\alpha$; moreover, it parametrizes isomorphism classes of line bundles with connection:

\begin{theorem}[quantization classes]
The set of Hermitian isomorphism classes of Hermitian line bundles with connection on a Riemannian manifold $M$ is given by the hypercohomology group $H^1(M,K)$ of the complex of sheaves $K$  \eqref{equ:2TKK}.
\end{theorem}

Since we are interested in quantizations for a given magnetic field, we will elaborate on isomorphism classes for fixed $L$ and $b$:

\begin{theorem}[quantization classes for fixed line bundle]
Let $(M,g,b)$ be a quantizable system with magnetic field and  $L$ a complex line bundle over  $M$ with $H^*(i)\left(c_1(L)\right)=[-\frac{1}{2\pi}b]$.
Then the set of equivalence classes of quantizations $(L,h,\nabla)$ of $(M,g,b)$ for fixed $(L,h)$ is given by $H^1(M,\R)/H^1(M,\Z)$.
\end{theorem}
\begin{proof} The set of Hermitian connections is parametrized by $\Omega^1(M)$ since two Hermitian connections differ by an imaginary 1-form $-\imath \eta$. 
Because $\curv(\nabla)=\curv(\nabla-\imath \eta)=\curv(\nabla)+d\eta$ we have $d\eta=0$, so $\eta=dk_\alpha$ for a suitable bundle atlas and $k_\alpha\in \check C^0(M,\underline{\R})$.  
Two quantizations $(L,h,\nabla)$ and $(L,h,\nabla'=\nabla+\imath \eta)$ are equivalent if and only if there is a Hermitian line bundle isomorphism with 
\[ \imath(a'_\alpha-a_\alpha) = f_\alpha^{-1}df_\alpha\] (see \eqref{equ:a'-a=fdf}). 
Therefore $\eta=a'-a=-\imath d\log f$. 
On the other hand, using the Bockstein homomorphism $j=\delta\circ\frac{\log}{2\pi\imath}:H^0(M,\underline{S^1}_M)\rightarrow H^1(M,\Z)$ one has
\[ g'=g\delta f=g \Leftrightarrow \delta f = 1 \Rightarrow j([f])\in H^1(M,\Z), \]
and such $f$ exist if and only if $\eta$ is integral.
So the sequence
\[ 0 \rightarrow H^1(M,\Z)\rightarrow H^1(M,\R)\rightarrow \Omega^{1,\text{closed}}(M)/\sim \rightarrow 0 \]
is exact; here two closed 1-forms $\eta_1,\eta_2$ are equivalent (``$\sim$'') if the connections $\nabla-\imath\eta_1$ and $\nabla-\imath\eta_2$ are equivalent.
\end{proof}

\begin{defi}[Jacobi torus]
$J(M):=H^1(M,\R)/H^1(M,\Z)$ is called the \emph{{Jacobi torus}} of $M$. The metric on $M$ induces a metric on $H^1(M,\R)$ and $H^1(M,\Z)$ via 
\[ (\eta,\omega):= \int_M \eta\wedge *\omega. \]
$J(M)$ carries the quotient topology.
\end{defi}

\begin{defi}[flat line bundle]
A line bundle is called \emph{flat} if there is a bundle atlas with locally constant transition functions.
\end{defi}

\begin{lemma}[classes of flat line bundles]\label{lemma:CFL}
The group (w.r.t.\ tensor product) of classes of flat line bundles on a manifold $M$ is isomorphic to the grouup $H^1(M,S^1)$.
\end{lemma}
\begin{proof} Flat line bundles are just locally constant line bundles. 
Thus a line bundle cocycle is a  \v{C}ech 1-cocycle with values in the locally constant $S^1$-valued functions. 
\v{C}ech coboundaries are exactly the isomorphisms of flat line bundles so that the set of classes of flat line bundles corresponds to the set of classes of \v{C}ech 1-cocycles. 
Finally, the cocycle of a tensor product is given by the product of the cycles of the factors.
\end{proof}

\begin{theorem}[quantization classes]\label{theorem:quantization classes}
For a Riemannian manifold $(M,g)$ with quantizable magnetic field $b$ the set of equivalence classes of quantizations $(L,h,\nabla)$  corresponds to $H^1(M,S^1)$.
\end{theorem}
\begin{proof} For a given choice $(L_1,h_1,\nabla_1)$ of a quantization every quantization $(L_2,h_2,\nabla_2)$ is -- modulo equivalence -- given by
\begin{align*}
(L_2,h_2,\nabla_2) &\simeq (L_1\otimes L_{12},h_1\otimes h_{12},\nabla_1\otimes\id_{L_{12}}+\id_{L_1}\otimes\nabla_{12}) \text{ with} \\
L_{12} &= L_1^*\otimes L_2,\\
h_{12} &= \overline{h_1}\otimes h_2,\\
\nabla_{12} &= \nabla_{1^*}\otimes\id_{L_2}+\id_{L_1^*}\otimes\nabla_2.
\end{align*}
Therefore the characterization of flat line bundle following lemma \ref{lemma:CFL} gives the set of quantization classes.
\end{proof}

\section{Connections}\label{sec:C}

First we will identify the Jacobi torus with the connected component of the unit in the group of one-dimensional unitary representations of the fundamental group of of $M$:
\begin{lemma}[Jacobi torus]\label{lemma:Jacobi torus}
\begin{equation} 
H^1(M,\R)/H^1(M,\Z)\simeq \left( \widehat{\pi_1(M)} \right)_0 \end{equation}
\end{lemma}
\begin{proof}
For every manifold $M$, $H:=H_1(M,\Z)$ is the abelization of $\Gamma:=\pi_1(M)$ so that $\hat H=\hat\Gamma$. 
As in \citep{KatSun:HCGCRS} we define the mapping
\begin{equation} \begin{split} \Omega^{1,\text{closed}}(M)\ni\omega &\mapsto \chi_\omega\in \hat H , \\
  \chi_\omega(\gamma) &:= \exp\left( 2\pi\imath\int_{c(\gamma)} \omega\right), \label{equ:chiomega}
		 \end{split}
\end{equation}
for a closed path $c(\gamma)$ representing the class $\gamma$. 
The integral does not depend on the choice of path since $\omega$ is closed. 
On exact forms, the integral over closed paths vanishes so that we obtain a well-defined mapping
\begin{align} H^1(M,\R)\ni[\omega] &\mapsto \chi_\omega\in\hat H.  \label{equ:chi[omega]} \end{align}
It is a homomorphism of groups because  $\chi_\omega(\gamma)\chi_{\omega'}(\gamma)=\chi_{\omega+\omega'}(\gamma)$.
The kernel consists of the (classes of) closed 1-forms $\omega$  for which $\int_c\omega$ is integral  for all closed paths $c$, i.e.\ just  (classes of) integral 1-forms.

\eqref{equ:chiomega} is continuous for every $\gamma$ and thus defines a continuous mapping into $\hat H$. 
Since $H^1(M,\R)$ is connected the range of \eqref{equ:chi[omega]} is connected, and it contains the trivial character as image of the zero class.
\end{proof}

\begin{lemma}[torsion torus]\label{lemma:torsion torus}
The isomorphism
\begin{align} 
\widehat{\pi_1(M)} &\simeq H^1(M,S^1) \\
\intertext{can be realized  geometrically by association of flat line bundles:}
\chi &\mapsto F_\chi = \tilde M\times_\chi\C
\end{align}
\end{lemma}
\begin{proof}
Equality follows from the universal coefficient theorem \cite[see  e.g.\ ][chapter 15]{BotTu:DFAT3}
\[ H^1(M,S^1) = \Hom(H_1(M,\Z),S^1)\oplus \operatorname{Ext}(H_0(M,\Z),S^1), \]
since $H_0(M,\Z)$ is free ($\Rightarrow \operatorname{Ext}(H_0(M,\Z),S^1)$ trivial) and $\pi_1(M)$ has the same one-dimensional representaions as its abelization $H_1(M,\Z)$.

By lemma \ref{lemma:CFL} $H^1(M,S^1)$ is the set of classes of flat line bundles with respect to ``flat equivalence''. 
On the other hand, flat vector bundles are just the vector bundles which are associated to a representation of the fundamental group.
Therefore, flat line bundles correspond to bundles associated to one-dimensional representations of the fundamental group:
\[ H^1(M,S^1)\simeq \{ \tilde M\times_\chi\C\mid \chi\in\widehat{\pi_1(M)} \}/\sim \]
On $\tilde M\times_\chi\C$ the natural flat connection is given by restriction of the canonical connection $d$ of the trivial bundle $\tilde M\times \C$.

On the other hand, given a flat line bundle one gets back the character $\chi$ as holonomy of the connections around closed paths:
For a flat connection on a complex line bundle $L$ parallel transport around a closed path depends only on the homotopy class of the path and therefore defines a unitary representation $\rho$ of $\pi_1(M)$.
Thus parallel transport gives a line bundle isomorphism $L\simeq \tilde M\times_\rho\C$. 
Since connection forms are invariant under flat equivalence the holonomy gives a well-defined mapping of $H^1(M,S^1)$ into $\widehat{\pi_1(M)}$ which obviously is inverse to the mapping  ``associating to $\tilde M$''.
\end{proof}

\begin{remark}[torsion torus]
By lemma \ref{lemma:Jacobi torus} the Jacobi torus is just $\left(\widehat{\pi_1(M)}\right)_0$. 
Decomposing $\Gamma$ into free (finitely generated) and  (finite) torsion parts one sees that characters in $\left(\widehat{\pi_1(M)}\right)_0$ are just the ones vanishing on the torsion part. 
The subsequence
\[
  0 \xrightarrow{}   H^1(M,\Z)  \xrightarrow{H^1(i)}   H^1(M,\R)  \xrightarrow{H^1(\exp2\pi\imath\cdot)}   H^1(M,S^1) \]
of the exact sequence \eqref{equ:exSequZRS1} shows that the Jacobi torus is embedded in $H^1(M,S^1)$  and does not contain torsion elements. 
Therefore $H^1(M,S^1)$ is the ``torsive version'' of the Jacobi torus, hence its name.
 \end{remark}

\section{Bloch theory on vector bundles}\label{sec:BTVB}
In this section we recall the basic elements of Bloch theory for periodic operators in the geometric context of  vector bundles. In the final section we will use it in the case of possibly non-trivial complex line bundles.
The standard reference for the theory of direct integrals is \citep{Dix:AOEHAN}, for Bloch theory in Euclidean space see \citep{ReeSim:AO}. 

Our general assumptions are: $X$ is an oriented smooth Riemannian manifold without boundary, $\Gamma$ a discrete abelian group acting on $X$ freely, isometrically, and properly discontinuously.
Furthermore, we assume the action to be cocompact in the sense that the quotient $M:=X/\Gamma$ is compact.

Next, let $E$ be a smooth Hermitian vector bundle over $X$.

\begin{defi}[periodic operator]\label{defi:periodic operator}
Assume there is an isometric  lift $\gamma_*$ of the action of $\gamma$ fom $X$ to $E$ in the following sense:
\begin{align}
\gamma_*:E_x\rightarrow E_{\gamma x}\text{ for }x\in X,\gamma\in\Gamma.
\end{align}
This defines an action $T_\gamma$ on the sections: For $s\in C^\infty_c(E)$ we define
\begin{align}
(T_\gamma s)(x) := \gamma_* s(\gamma^{-1}x)\text{ for }x\in X,\gamma\in\Gamma.
\end{align}
$(T_\gamma)_{\gamma\in\Gamma}$ induces a unitary representation of $\Gamma$ in $L^2(E)$ since $\gamma_*$ acts isometrically and $T_\gamma^*=(T_\gamma)^{-1}$.

A differential operator $D$ on ${\mathcal D}(D):=C^\infty_c(E)$ is called periodic if, on $\mathcal{D}(D)$, we have: 
\begin{equation} \forall{\gamma\in\Gamma}: [T_\gamma,D]=0 \end{equation} 
\end{defi}

\begin{lemma}[associated bundle]
$E$ is the lift  $\pi^*E'$ of a Hermitian vector bundle $E'$ over $M$ by the projection $\pi:X\rightarrow M$. $E$ and $X$ are $\Gamma$-principal fiber bundles over $E'$ resp.\ $M$.

To every $\Gamma$-principal fiber bundle and every character $\chi\in\hat\Gamma$
we associate a line bundle. 
This gives the relations depicted in the following diagram
  (``$\rightsquigarrow$'' denotes association of line bundles.):

\begin{equation*}
\begin{CD}
       @.   \C^N @.   \C^N  @.  @. @. \C^N @. \C^N\\
   @.       @VVV      @VVV   @. @. @VVV @VVV  \\
\Gamma &\hookllongrightarrow& E    @>\textstyle\pi_*>> E'    @. \quad\rightsquigarrow\quad @. \C &\hookllongrightarrow& E_\chi    @>>> E' \\
   @.       @VV\textstyle\pi^EV      @VV\textstyle\pi^{E'}V @.                          @. @VVV @VVV \\
\Gamma &\hookllongrightarrow& X    @>\textstyle\pi>> M   @. \quad\rightsquigarrow\quad @. \C &\hookllongrightarrow& F_\chi    @>>> M 
\end{CD}
\end{equation*}
\begin{center}principal fiber bundles and associated line bundles\end{center}\label{fig:PAV}

In this situation we have $E_\chi\simeq E'\otimes F_\chi$.
\end{lemma}
\begin{proof}
$E$ is a $\Gamma$-principal fiber bundle, so we can use the lifted  $\Gamma$-action  to define $E':=E/\Gamma$.
Since this action is a lift of the $\Gamma$-action on $X$, $E'$ has a natural structure of a vector bundle over $M$. If $\pi^{E'}:E'\rightarrow M$ is the bundle projection of $E'$, then the pull back by $\pi$  is defined as
\begin{align*}
\pi^* E' &= X\times_\pi E'=\{ (x,e)\in X\times E'\mid \pi(x)=\pi^{E'}(e) \}.
\end{align*}
If $\pi^{E}:E\rightarrow X$ is the bundle projection of $E$ and $\pi_*:E\rightarrow E'$ is the quotient map, then we get a bundle isomorphism $E\rightarrow \pi^* E'$  by
\begin{align*}
E\ni e \mapsto (\pi^E(e),\pi_*(e)) \in \pi^* E'.
\end{align*}
Therefore, in this representation the lift $\gamma_*$ of $\gamma$ acts on $(x,e)\in\pi^*E'$ as $\gamma_*(x,e)=(\gamma x,e)$.

Sections into an associated bundle $P\times_\rho V$ are just those sections of the bundle $P\times V$ which have the appropriate transormation property. 
By construction, $E_\chi$ is a complex line bundle over $E'$, but from $E$ it inherits the vector bundle structure, so its sections fulfill:
\begin{equation} C^\infty(E_\chi) \simeq C^\infty(E)^{\Gamma,\chi}=\{s\in C^\infty(E)\mid \forall{\gamma\in\Gamma}:\gamma^*s=\chi(\gamma)s \} \label{equ:Echi=EGchi} \end{equation}
An analogous equation holds for the line bundle $F_\chi$ over $M$.
Finally, \eqref{equ:Echi=EGchi} shows
\begin{align*} E_\chi &= E\times_\chi\C \\
 &= (\pi^* E')\times_\chi\C \\
 &= (X\times_\pi E')\times_\chi\C \\
 &\simeq E'\otimes (X\times_\chi\C) \\
 &= E'\otimes F_\chi.
\end{align*}
Here, all equalities are immediate from the definitions, besides the last but one, which may be seen as follows:
\begin{align*}
(X\times_\pi E')\times_\chi\C &= (X\times_\pi E'\times\C)/\Gamma
\intertext{with the $\Gamma$-action}
\gamma(x,e,z) &= (\gamma x,e,\chi(\gamma)z),
\intertext{whereas}
E'\otimes (X\times_\chi\C) &=E'\otimes ((X\times\C)/\Gamma)
\intertext{with the $\Gamma$-action}
\gamma(x,z) &= (\gamma x,\chi(\gamma) z).
\end{align*}
So, both bundles are quotients of isomorphic bundles with respect to the same $\Gamma$-action.
\end{proof}

Next we want to decompose the Hilbert space $L^2(E)$ of square-integrable sections of $E$ into a direct integral over the character space $\hat\Gamma$. 
On $\hat\Gamma$ we use the Haar measure. 
From the theory of representations of locally compact groups we need the following character relations for abelian discrete $\Gamma$, i.e.\ for abelian, compact $\hat\Gamma$ \cite[see e.g.\ ][ \S 1.5]{Rud:FAG}:

\begin{lemma}[character relations]
For $\gamma\in\Gamma$ 
\begin{equation} \int_{\hat\Gamma}\chi(\gamma)\,d\chi = \left\{ \begin{array}{l} 1, \quad \gamma=e, \\ 0, \quad \gamma\neq e. \end{array} \right. \label{equ:Charakter}\end{equation}
For $\chi,\chi'\in\hat\Gamma$ 
\begin{equation} \sum_{\gamma\in\Gamma} \bar\chi(\gamma)\chi'(\gamma) = \delta(\chi-\chi') \label{equ:Charakter'}\end{equation}
in distributional sense, i.e.\ for $f\in C(\hat\Gamma)$ 
\[ \sum_{\gamma\in\Gamma} \int_{\hat\Gamma}\bar\chi(\gamma)\chi'(\gamma)f(\chi)\,d\chi = f(\chi'). \]
\end{lemma}

We define for every character $\chi\in\hat\Gamma$ a mapping $\Phi_\chi:C^\infty_c(E)\ni s\mapsto \tilde s_\chi\in C^\infty(E)$ by
\begin{equation} \tilde s_\chi(x) := \sum_{\gamma\in\Gamma} \chi(\gamma)\gamma_*s(\gamma^{-1}x). \label{equ:schi} \end{equation}
Since
\begin{align*}
 \tilde s_\chi(\gamma' x) &= \sum_{\gamma\in\Gamma}\chi(\gamma)\gamma_*s(\gamma^{-1}\gamma'x) \\
 &= \sum_{\gamma\in\Gamma}\chi(\gamma'\gamma'{}^{-1}\gamma)(\gamma'\gamma'{}^{-1}\gamma)_*s\left((\gamma'{}^{-1}\gamma)^{-1}x\right) \\
 &= \chi(\gamma')\gamma'_* \tilde s_\chi(x)
\end{align*}
we have
\[ \tilde s_\chi\in C^\infty(E)^{\Gamma,\chi}=\{r\in C^\infty(E)\mid \forall_{\gamma\in\Gamma} T_\gamma r = \chi(\gamma) r\} \] 
which defines a section $s_\chi\in C^\infty(E_\chi)$. 

Let $\mathcal D$ be a fundamental domain for the $\Gamma$-action, i.e.\ an open subset of $X$ such that $\bigcup_{\gamma\in\Gamma}{\gamma\mathcal D}=X$ up to a set of measure 0 and $\gamma\mathcal D\cap\mathcal D=\emptyset$ for $\gamma\not=e$. Then
\begin{align*}
 \int_{\hat\Gamma} \|s_\chi\|^2_{L^2(E_\chi)}d\chi &= \int_{\hat\Gamma} \int_{\mathcal{D}} |\tilde s_\chi(x) |^2 dx\,d\chi \\
 &= \int_{\mathcal{D}} \int_{\hat\Gamma} \sum_{\gamma_1,\gamma_2\in\Gamma} \chi(\gamma_1^{-1}\gamma_2)\langle {\gamma_1}_*s(\gamma_1^{-1}x)\mid{\gamma_2}_*s(\gamma_2^{-1}x)\rangle_{E} d\chi\, dx \\
 &= \int_{\mathcal{D}} \sum_{\gamma\in\Gamma} |s(\gamma^{-1}x)|^2 dx \\
 &= \|s\|^2_{L^2(E)}.
\end{align*}
On the one hand, this shows that we can define a measurable structure on $\prod_{\chi\in\hat\Gamma}L^2(E_\chi)$ by choosing a sequence in $C^\infty_c(E)$ which is total in  $L^2(E)$.
On the other hand, we can see that the direct integral $\int^\oplus_{\hat\Gamma}L^2(E_\chi)\,d\chi$ is isomorphic to $L^2(E)$ via the isometry
$\Phi$, whose inverse is given by
\[ \Phi^*\colon(s_\chi)_{\chi\in\hat\Gamma} \mapsto \int_{\hat\Gamma} \tilde s_\chi(x)\,d\chi, \]
as is easily seen from the character relations \eqref{equ:Charakter} and \eqref{equ:Charakter'}.

This shows

\begin{lemma}[direct integral]
 The mapping defined by \eqref{equ:schi} can be extented continuously to a unitary
\begin{equation} \Phi: L^2(E) \rightarrow \int^\oplus_{\hat\Gamma} L^2(E_\chi)\,d\chi. \end{equation}
\end{lemma}

For the direct integral of Hilbert spaces $H=\int^\oplus_{\hat\Gamma}H_\chi d\chi$ 
the set of decomposable bounded operators
$L^\infty(\hat\Gamma,{\mathcal L}(H))$  is given by the commutant
$(L^\infty(\hat\Gamma,\C))'$ in ${\mathcal L}(H)$.
Since commutants are weakly closed and $C(\hat\Gamma,\C)$ is weakly dense in $L^\infty(\hat\Gamma,\C)$ one has $(L^\infty(\hat\Gamma,\C))'=(C(\hat\Gamma,\C))'$.
Therefore, in order to determine the decomposable  operators one has to determine the action of $C(\hat\Gamma)$ on $L^2(E)$. This is easily done using the explicit form of $\Phi$:

\begin{propo}[$C(\protect\hat{\Gamma})$-action]
$f\in C(\hat\Gamma)$ acts on $s\in C^\infty_c(E)$ by
\begin{align}
 M_fs &:= \Phi^*f\Phi s, \\
\intertext{and one has}
 (M_fs)(x) &= \sum_{\gamma\in\Gamma} \hat f(\gamma^{-1})T_\gamma s(x),\text{ where} \label{equ:fAktion} \\
 \hat f(\gamma) &:= \int_{\hat\Gamma}f(\chi)\bar\chi(\gamma)\,d\chi \label{equ:fhat=intf}
\end{align}
is the Fourier transform of $f$. 
$M_f$ is a bounded operator with norm $\|f\|_\infty$.
\end{propo}
\begin{proof}
For $x\in X$ one has:
\begin{align*}
 (M_fs)(x) &= (\Phi^*f\Phi s)(x) \\
 &= \int_{\hat\Gamma} (f\Phi s)_\chi(x)\,d\chi \\
 &= \int_{\hat\Gamma} f(\chi) \sum_{\gamma\in\Gamma}\chi(\gamma)\gamma_* s(\gamma^{-1}x)\,d\chi \\
 &= \sum_{\gamma\in\Gamma} \hat f(\gamma^{-1})\gamma_* s(\gamma^{-1}x)
\end{align*} 
Since $f$ is a multiplication operator in each fiber it has fiberwise norm $\|f\|_\infty$, and so have $f$ and $M_f=\Phi^*f\Phi$.
\end{proof}

\begin{corol}[decomposable operators]
Conjugation by $\Phi$ defines an isomorphism between decomposable bounded operators on $\int^\oplus_{\hat\Gamma}L^2(E_\chi)\,d\chi$ and  $\Gamma$-periodic bounded operators on $L^2(E)$.
\end{corol}
\begin{proof} \item[]
\begin{description}
\item[``$\Rightarrow$''] A decomposable operator commutes with the $C(\hat\Gamma)$-action, especially with $f_\gamma\in C(\hat\Gamma)$ which is defined by
\[ \hat f_\gamma(\gamma') := \begin{cases} 1,& \text{if }\gamma=\gamma', \\ 0& \text{else.}  \end{cases} \]
By~\eqref{equ:fAktion} commuting with $f_\gamma$ is equivalent to commuting with $\gamma$.
\item[``$\Leftarrow$'']  To commute with the $\Gamma$-action means to commute with all $f_\gamma$ for $\gamma\in\Gamma$. Because of
\[ f_\gamma(\chi) = \chi(\gamma) \]
the $f_\gamma$ are just the characters $\widehat{\hat\Gamma}$ of the compact group $\hat\Gamma$, and by the Peter-Weyl theorem (or simpler: by the Stone-Weierstra\ss{} theorem) they are dense in $C(\hat\Gamma)$. 
Since the operator norm  of $M_f$ and the  supremum norm of $f$ coincide the commutation relation follows for all $f\in C(\hat\Gamma)$ by continuity.
  \end{description}
\end{proof}

An unbounded operator is decomposable if and only if its (bounded) resolvent is decomposable.
For a periodic symmetric elliptic operator $D$ we have a domain of definition ${\mathcal D}(D)=C^\infty_c(X)$ on which $D$ is essentially self-adjoint.
This domain is invariant for $D$ as well as for the $\Gamma$-action,
and  one has $[D,\gamma]=0$ for all $\gamma\in\Gamma$. 
Thus all bounded functions of $D$ commute with the $\Gamma$-action, and one has:

\begin{theorem}[decomposition of periodic operators]\label{theorem:DPO}
The closure $\bar D$ of every periodic symmetric elliptic operator $D$ is decomposable with respect to the direct integral $\int^\oplus_{\hat\Gamma}L^2(E_\chi)\,d\chi$.
A core for the domain of $\bar D_\chi$ is given by $C^\infty(E_\chi)$, and the action of  $D_\chi$ on $C^\infty(E_\chi)\simeq C^\infty(E)^{\Gamma,\chi}$ is just the action of $D$ as differential operator on $C^\infty(E)^{\Gamma,\chi}$. 
We have $\bar D_\chi=\overline{D_\chi}$, where $D_\chi:=D|_{C^\infty(E)^{\Gamma,\chi}}$, and the closures are to be taken as operators in $L^2(E_\chi)$.
\end{theorem}
\begin{proof}
Given the remark above we have shown the decomposability already.

 $C^\infty_c(X)$ is a core for $\bar D$, its image under $\Phi_\chi$ is contained in $C^\infty(E)^{\Gamma,\chi}$ and is a core for $\bar D_\chi$, sinece $\Phi$ is an isometry.
On this domain  \eqref{equ:schi} gives the action of $\bar D_\chi$ as asserted in the theorem.
Since $D_\chi$ is a symmetric elliptic operator on the compact manifold $M$ it is essentially self-adjoint. $\bar D_\chi$ is a fiber of $\bar D$ \citep[which is self-adjoint by, e.g.,][]{Ati:EODGNA} and therefore self-adjoint, thus both define the same unique self-adjoint extension $\overline{D_\chi}$ of $D_\chi$. 
\end{proof}

\section{Periodic magnetic fields}\label{sec:PMF}
From now on we assume the existence of a free isometric properly discontinuous action 
of a discrete group $\Gamma$ on the Riemannian manifold $X$.
We assume the action to be cocompact in the sense that the quotient manifold $M:=\Gamma\backslash X$ is compact. 
Furthermore, let $b\in\Omega^2(X)$ be a quantizable periodic magnetic field so that
\begin{align*}
 db &= 0, \\
 \left[\frac{1}{2\pi}b\right] &\in H^*(i)\left( H^2(X,\Z) \right)\subset H^2(X,\R), \\
 b &= \pi^*b_M\text{ for a }b_M\in \Omega^2(M),\\
\intertext{where}
 \pi &: X\rightarrow M
\end{align*}
is the projection. 
The main point is that integrality of $\frac{1}{2\pi}b_M$ is not automatic: 
For a two-dimensional manifold $X$integrality of $\frac{1}{2\pi}b_M$ means integrality of the magnetic flux $\int_Mb_M$ through one elementary lattice cell,
whereas $\frac{1}{2\pi}b$ is integral automatically if e.g.\ $H^2(X,\Z)=0$.

\begin{theorem}[periodic magnetic Schr\"odinger operator]\label{theorem:periodic magnetic Schroedinger operator}
If $\frac{1}{2\pi}b_M$ is integral then there exists a quantization $(L,h,\nabla)$ on $X$ such that the corresponding magnetic Schr\"odinger operator $H^{L,\nabla}$ is $\Gamma$-periodic.
\end{theorem}
\begin{proof}
If $\frac{1}{2\pi}b_M$ is integral there is a Hermitian line bundle $L'$ over $M$ with connection $\nabla'$ by theorem \ref{theorem:EQ}, so that $\curv(\nabla')=b_M$. 
$L'$ and $\nabla'$ can be pulled back via $\pi$ from $M$ to $X$, giving a line bundle $L=\pi^*L'$ over $X$ with connection $\nabla$ and curvature $b$. 

The $\Gamma$-action on $X$ induces a $\Gamma$-action on $L$: 
Let $\alpha_\gamma:X\rightarrow X$ be the action of $\gamma\in\Gamma$ on $X$. Then $\pi\circ\alpha_\gamma=\pi$, and therefore
\[ \alpha_\gamma^*L=\alpha_\gamma^*\pi^*L'=\pi^*L'=L. \]
Thus, $L=\{(x,l)\in X\times L'\mid l\in L'_{\pi(x)}\}$ carries a natural $\Gamma$-action by acting on the first component, using the action on $X$.
 
Since $\nabla$ is lifted by $\pi$ it is automatically $\Gamma$-periodic: because of the proper discontinuity of the $\Gamma$-action every finite covering of $M$ by open sets induces a locally finite covering of $X$ by $\Gamma$-invariant open sets, and the connection forms of $M$ can be pulled back to periodic forms on $X$.
\end{proof}

Usually one adds a smooth, periodic function $V$ (the ``electric potential'') to get the full Schr\"odinger operator.
The resulting operator is periodic and elliptic, therefore we can apply appropriate analytic methods. 
Especially,  $H^{L,\nabla}$ is essentially self-adjoint.
 If $\Gamma$ is abelian we have the Bloch decomposition:

\begin{theorem}[direct integral]\label{theorem:direct integral}
For abelian $\Gamma$ the fibers of  $\overline{H^{L,\nabla}}$ are given by
\begin{align}
\left(\overline{H^{L,\nabla}}\right)_\chi &= \overline{H^{L,\nabla}_\chi}\text{ with}\label{equ:dIa}\\
 {\mathcal D}\left( H^{L,\nabla}_\chi \right) &= C^\infty(L_\chi)=C^\infty(L)^{\Gamma,\chi},\\
 L_\chi &= L'\otimes   F_\chi,\\
H^{L,\nabla}_\chi &= H^{L,\nabla}|_{C^\infty(L)^{\Gamma,\chi}} \label{equ:dIo} \\
 &= H^{L_\chi,\nabla_\chi}, \label{equ:dIh}
\intertext{where}
 \nabla_\chi &= (\nabla'\otimes\id+\id\otimes d)|_{C^\infty(L)^{\Gamma,\chi}} \label{equ:dId}
\end{align}
In other words: every fiber of the magnetic Schr\"odinger operator $H^{L,\nabla}$ is a magnetic Schr\"odinger operator of type $H^{L_\chi,\nabla_\chi}$.
\end{theorem}
\begin{proof}
By \ref{theorem:DPO}  $H^{L,\nabla}$ is decomposable, and equations \eqref{equ:dIa}--\eqref{equ:dIo} follow immediately. 
Equation \eqref{equ:dIh} follows for $\chi=1$ from Leibniz's rule for connections since $F_\chi=M\times\C$ in this case. 
For all $\chi$  \eqref{equ:dId} defines, as we have seen in the proof of lemma \ref{lemma:torsion torus}, a connection for the quantization class characterized by $\chi$ following theorem \ref{theorem:quantization classes}.
Moreover, the explicit form shows that
\[ \nabla|_{C^\infty(L)^{\Gamma,\chi}}=\nabla_\chi \]
since $\nabla_\chi$ does not depend on $\chi$ explicitly. 
Therefore
\[ H^{L,\nabla}|_{C^\infty(L)^{\Gamma,\chi}} = H^{L_\chi,\nabla_\chi}, \]
and the proof is completed by \eqref{equ:dIo}.
\end{proof}

\begin{corol}[Bloch theory and quantization]\label{corol:Bloch theory and quantization}
Let $(M,g,b)$ be a quantizable system with magnetic field over a compact manifold $M$. 
Then the corresponding system $(\tilde M,\tilde g,\tilde b)$ is (up to equivalence) uniquely quantizable on the universal covering space.
Moreover, if $\Gamma:=\pi_1(M)$ is abelian then the magnetic Schr\"odinger operator $H^{\tilde L,\tilde\nabla}$ on $L^2(\tilde M)$ is decomposable over $\hat\Gamma$, and the fibers occuring are just the equivalence classes of quantizations of $(M,g,b)$:
\begin{align} \overline{H^{\tilde L,\tilde\nabla}} &= \int_{\widehat{\pi_1(M)}}^\oplus \overline{H^{L_\chi,\nabla_\chi}}\,d\chi \\
 \text{``unique quantization above''} &= \text{``sum over all quantizations downstairs''} \notag
\end{align}
\end{corol}
\begin{proof}
The system $(\tilde M,\tilde g,\tilde b)$ is periodic and quantizable by construction. 
Since obviously  $H^1(\tilde M,S^1)=\{1\}$ the quantization is unique up to equivalence.
For abelian $\Gamma$ we can apply theorem \ref{theorem:direct integral} from which, together with theorem \ref{theorem:quantization classes}, we get the conclusion.
\end{proof}

\begin{remark}[non-abelian fundamental group]
Even if $\pi_1(M)$ is non-abelian one may choose a homology covering space $X$ of $M$ such that the covering group is abelian ($H_1(M,\Z)$) and $\pi_1(X)$ is finite (the torsion part). 
Now there is a finite number of classes of quantizations on $X$, the set of classes of quantizations on $M$ has a finite number of components. 
Bloch analyzing a quantization on $X$ (with respect to the abelian group $H_1(M,\Z)$) now gives all quantizations on $M$ belonging to one component of $\widehat{\pi_1(M)}$, generalizing the previous corollary. 
Note that this does not yet allow to decompose the periodic operators on $X$ with respect to the full non-abelian group $\widehat{\pi_1(M)}$.
\end{remark}

\begin{remark}[non-commutative Bloch theory]
Given the previous remark it is natural to try to decompose the periodic operators with respect to a non-abelian group. 
This may be a group of translations or a variation thereof, the so-called magnetic translations. 
In any case there is no good character group $\hat\Gamma$ any more which would allow for the Fourier transform which one uses in the abelian case: the set of irreducible representations lacks the group structure, the set of one-dimensional representations is to small to describe the whole group (or the group including the magnetic gauge).
But the space $C(\hat\Gamma)$ of continuous functions on $\hat\Gamma$ continues to exist in the non-abelian case in the form of the reduced group \CS-algebra of $\Gamma$. 
This may be viewed as a non-commutative topological space or --- after recognizing additional natural structures on it --- as a non-commutative Riemannian manifold in the sense of \citet{Con:NG}.

Depending on the different aims (index and K-theory, transport properties and quantum Hall effect, spectral theory) and assumptions (free group actions, transitive projective actions, free projective actions) this observation has been used in different manners. 
The last mentioned case relates most to the subject of this paper, and we refer --- slightly biased --- to \citep{Gru:NB, Gru:NBTO} and the references therein.

\end{remark}

\ifx\undefined\allcaps\def\allcaps#1{#1}\fi
  \ifx\undefined\nop\newcommand{\nop}[1]{}\fi
  \ifx\undefined\single\newcommand{\single}[1]{#1}\fi
  \ifx\undefined\SwapArgs\newcommand{\SwapArgs}[2]{#2#1}\fi
  \ifx\undefined\translationof\newcommand{\translationof}{English Translation
  of }\fi \ifx\undefined\submitted\newcommand{\submitted}{Submitted }\fi
  \ifx\undefined\submittedto\newcommand{\submittedto}{Submitted to }\fi
  \ifx\undefined\privcomm\newcommand{\privcomm}{Private communication}\fi
  \providecommand{\inpreparation}{In preparation}
  \providecommand{\toappearin}{To appear in }

\end{document}